# Variaciones geomagnéticas diarias debido al efecto lunar: estudio basado en datos del observatorio geomagnético de Huancayo

Domingo Rosales[*]
Erick Vidal[**]

**Abstract**

The Moon apparently seems to have not appreciable effect in the geomagnetic monthly period, but Keil (1839), Sabine (1853), Broun (1874) and many others have shown a lunisolar daily variation. It is known that solar (S) and lunar (L) variation have seasonal variations. Seasonal changes of S, particularly on quiet days, have been studied in greater detail than the seasonal changes for L. The purpose of this paper is to describe in more detail the effect of the variation of L in the geomagnetic equator, in the absence of strong perturbations selecting conveniently geomagnetic data of Huancayo observatory (hourly mean) from January 1, 2008 to December 31, 2009, period which is longer in range of minimum solar activity of the last 160 years (end of solar cycle 23 and start of solar cycle 24). The spectral analysis by wavelet and Fourier allows us to identify the main contribution of S diurnal and semidiurnal variations and its sidebands, besides the effect of the seasonal variation. In order to observe the variation L is subtracted the variation S together with its sidebands, allowing observed by Fourier spectral analysis and wavelet as the main contributions to the variation L, similar effect were described by Chapman and Miller (1940).

Key words: *Solar and lunar daily geomagnetic variations, solar variations, lunar variations, spectral analysis, wavelets, fast Fourier transform* .

[**] Observatorio geomagnético de Huancayo, Instituto Geofísico del Perú, domingo.igp@gmail.com
[***] Observatorio geomagnético de Huancayo, Instituto Geofísico del Perú, erick.vidal@igp.gob.pe

**Resumen**

La Luna aparentemente parece no tener un efecto geomagnético apreciable de periodo mensual, pero Keil (1839), Sabine (1853), Broun (1874) y muchos otros han demostrado que existe una variación lunisolar diaria. Se sabe que la variación solar (S) y lunar (L) tienen variaciones estacionales. Los cambios estacionales de S, particularmente en días quietos, han sido estudiados en mucho más detalle que los cambios estacionales de L. El propósito del presente trabajo es describir con mayor detalle el efecto de la variación L en la zona del ecuador geomagnético, en ausencia de fuertes perturbaciones por el que se ha seleccionado convenientemente los datos del observatorio geomagnético de Huancayo (promedios horarios) del 1 de Enero del 2008 al 31 de Diciembre del 2009, periodo que se encuentra dentro del rango más prolongado de mínima actividad solar de los últimos 160 años (fin del ciclo solar 23 e inicio del ciclo solar 24).

El análisis espectral por wavelets y Fourier nos permite identificar la variación S compuesto por sus armónicas, siendo $S_1$ y $S_2$ la principales contribución que corresponden a la variación diurna y semidiurna, así como de sus bandas laterales, además del efecto de la variación estacional. Con la finalidad de poder observar la variación L, es sustraído la variación S juntamente con sus bandas laterales, permitiendo observar mediante el análisis espectral por wavelets $L_1$ y $L_2$ como las principales contribuciones en la variación L, además del efecto estacional similar a lo descrito por Chapman y Miller (1940).

Palabras clave: *variaciones geomagnéticas solar y lunar diaria, variación solar, variación lunar, análisis espectral, wavelets, transformada rápida de Fourier.*

**Introducción**

El campo geomagnético es en su mayor parte de *origen interno*. Sin embargo existe también un campo de *origen externo*, generado principalmente por la actividad solar. Una de las principales manifestaciones temporales del campo geomagnético de origen interno es *la "variación secular"* que sólo es apreciable en periodos largos de tiempo (mayor a dos años), en tanto que la *"variación diurna"* es una de las principales manifestaciones del campo geomagnético de origen externo, las cuales son observables en periodos mas cortos (de 3 a 25 horas aproximadamente).

El estudio de las variaciones diurnas y su naturaleza no es simple. Son muchos los factores que influyen y llevan a diferentes tipos de resultados, dependiendo del tipo de datos usados y del método de análisis aplicado.

Los principales factores que influyen en la variación diurna son; de *efecto solar*: actividad solar y rotación solar;  de *efecto terrestre*: traslación de la Tierra alrededor del Sol, rotación de la Tierra, corrientes eléctricas ionosféricas y terrestres; y de *efecto lunar*: corrientes eléctricas ionosféricas, oceánicas y terrestres. Todos estos factores complican el análisis en las variaciones diurnas.

En las variaciones diurnas se pueden distinguir dos tipos de variaciones *"periódicas"* y *"no periódicas"*. Las variaciones periódicas son mejor observados en los días tranquilos, es decir en aquellos días en que la perturbación de la actividad solar es mínima. En el análisis de las variaciones periódicas diurnas nos permite distinguir hasta tres tipos de variaciones principales, debido al efecto solar (S), debido al efecto lunar (L), y debido al efecto de la rotación solar (R) (De Meyer, 2003).

La variación S tiene un periodo predominante diurna (24 horas), mientras que la variación L tiene un periodo predominante semidiurna (12.42 horas) . La variación L debido a su pequeña amplitud, en relación con la producida por el Sol, no se aprecia directamente en los magnetogramas, sino que se debe deducir a través de un análisis armónico de varios meses de observaciones, lo usual para este tipo de análisis es utilizar todos los días del mes exceptuando los días perturbados.

En un análisis espectral la Luna aparentemente parece no tener un efecto geomagnético apreciable de periodo mensual (Střeštĭk, 1998), pero en el año 1841, M. Kreil usando los datos de Declinación de Praga anunció en la Sociedad de Ciencias de Bohemia, que descubrió la influencia lunar en la variación del campo geomagnético, luego de dicho descubrimiento E. Sabine, en 1853 y algunos otros científicos estudiaron y confirmaron este fenómeno para las tres componentes geomagnéticas. Van Bemmelin en 1912 fue el primero que trato la variación geomagnética lunar como un fenómeno global y analizo sus datos usando el método de los harmónicos esféricos desarrollados por Schuster en 1889. Años más tarde S. Chapman, entre 1913 a 1917 hizo un estudio más detallado de la variación geomagnética lunar en estaciones individuales y expandió los valores de los harmónicos esféricos obtenidos. En 1936 Bartels J., descubrió el marcado dominio de la amplitud semidiurna de la componente horizontal de la variación lunar, basada en datos del observatorio de Huancayo, ubicado dentro de la zona del ecuador geomagnético. El mismo tipo de acrecentamiento fue reportado por los observatorios de Kodaikanal 1958, Ibadan 1959 y Addis Ababa 1960, por lo que en 1962 S. Matsushita, propuso que ello seria debido a un efecto del *electrochorro* lunar en la zona ecuatorial magnética (Matsushita, 1967). Chapman y Lindzen en 1970 publicaron un tratado sobre mareas atmosféricas. En 1981 Winch reviso las técnicas de análisis lunar y estudio las variaciones lunares durante los años 1964 y 1965. Matsushita y Xu entre 1983 y 1984 reexaminaron los datos de la variación lunar de los primero estudios del Año Geofísico Internacional (Cueto, 2001).

El cambio estacional de S ha sido estudiado por muchos investigadores en mucho más detalle que el cambio estacional de L, particularmente para los días quietos. Por lo que el propósito del presente trabajo es estudiar las variaciones del campo geomagnético en el observatorio de Huancayo, debido a las influencia de la Luna en ausencia de fuertes perturbaciones.

**Orígenes del las corrientes y campos de la variación lunar**

Teóricamente las variaciones geomagnéticas lunares son debido a un proceso de circulación de corrientes eléctricas localizados en la región dinamo de la ionosfera aproximadamente a 100 Km. de altitud y de corrientes secundarias, las cuales son inducidos por las corrientes ionosféricas, localizados en la corteza terrestre conductora (Cueto, 2001). Las corrientes ionosféricas responsable de las mareas geomagnéticas lunares son generadas por la marea atmosférica lunar que es gravitacionalmente forzado por la Luna, principalmente en la baja atmósfera, y que se propaga hacia la alta atmósfera. En la ionosfera la intensidad de las corrientes ionosféricas depende de la velocidad de los vientos y de la conductividad de la ionosfera. La conductividad ionosférica depende básicamente de la densidad de electrones, y de la gran dependencia del tiempo solar local. De esta manera la marea geomagnética lunar cambia con el tiempo solar así como con la edad lunar (Yamazaki et al, 2012).

El efecto gravitacional de la Luna no solo actúa sobre la atmósfera terrestre, sino que también sobre los océanos y la parte sólida terrestre (esta ultima es poco conocido pero que contribuye con una marea gravitacional terrestre de +/- 20 cm. y puede llegar hasta 30 cm. Los movimientos de marea de los océanos también generan corrientes eléctricas de dinamo en el agua salada conductora, que también contribuyen en la variación del campo geomagnético (Malin, 1970).

Los primeros estudios integrales de la influencia de la Luna sobre el campo geomagnético fueron realizados por Bartels y Jonhston en 1940, con datos de la intensidad total H del observatorio de Huancayo (periodo de Marzo 1922 a Octubre 1939), son analizados varios métodos de estudio de la variación L, de acuerdo a sus resultados determinaron que en los meses de Noviembre a Marzo L es mayor que el resto del año, así mismo L se incrementa proporcionalmente al numero de manchas solares, pero que alrededor de Junio L es mínimo, también  determinaron que las variaciones S y L son independientes uno del otro (Bartels and Johnston, 1940a, 1940b).

**Datos**

Debido a su ubicación estratégica del observatorio de Huancayo (ubicado  en el ecuador geomagnético, con coordenadas geográficas: latitud 12º 02' 28.69'' Sur, longitud 75º 19' 14.11'' Oeste, altitud 3314 metros sobre el nivel del mar, y Dip[1] al 2009.5 de 0.56º), se ha seleccionado convenientemente los promedios horarios de la intensidad horizontal (H) del observatorio de Huancayo, del 1 de Enero del 2008 al 31 de Diciembre del 2009, incluyendo 731 ondas lunares ò 1462 ondas semilunares.

La resolución de la frecuencia del espectro calculado es adecuada para determinar cualquier periodo asociado con la Luna. Los años 2008 y 2009 es el periodo que se encuentra dentro del rango más prolongado de mínima actividad solar de los últimos 160 años, correspondiente al mínimo de actividad solar del ciclo 23 e inicio del ciclo solar 24, registrándose un total de 780 días sin manchas solares, por lo que la influencia de la actividad solar en la Tierra es mínimo y esto es observado en las condiciones tranquilas del campo magnético terrestre. Lo que permite contar con una serie de datos uniformemente equiespaciados temporalmente, requisito fundamental para poder aplicar adecuadamente el análisis espectral mediante wavelets y análisis de Fourier. A los datos seleccionados se sustrae el efecto de la variación secular quedando el residual para ser analizado (figura 1a).

**Método de Análisis**

Las variaciones L no son observables a simple vista en los magnetogramas ya que están enmascaradas por las variaciones S, que son mucho mayores. Por lo que se hace necesario tener varios meses de registro para aplicar el análisis espectral para identificar la contribución lunar.

Para la determinación de las variaciones geomagnéticas debido a la acción de la Luna, el método clásico usado es el desarrollado por Chapman y Miller (Chapman and Miller, 1940) el cual es descrito detalladamente por Malin y Chapman (Malin and Chapman, 1970), que consiste en agrupar los datos en meses de Lloyd[2] que corresponden a los solsticios de verano e invierno y a los dos equinoccios, analizándose el efecto lunar en cada grupo. Esta técnica generalmente es usado para la determinación de los primeros cuatro armónicos de la variación lunar, siendo la variación semidiurna $L_2$ la mas importante, seguido de la variación diurna $L_1$. Bajo esta metodología Chapman y Bartels (1940) concluyeron que la variación estacional de L es mucho mayor que la de Sq[3], mientras que Matsushita y Maeda (1965) determinaron que ambas variaciones eran similares (Matsushita and Maeda, 1965).

El presente análisis para las variaciones S y L es una extensión sobre el método tradicional de Chapman-Miller considerándose como una combinación lineal de las variaciones S y L mediante la relación $H(t) = S(t) + L(t)$. A los datos de la componente H sustraído la variación secular se aplica el análisis de wavelets y la transformada rápida de Fourier (FFT) en alta resolución con la finalidad de identificar las frecuencias fundamentales y su estructura fina (bandas laterales) de S. El análisis espectral por wavelets (figura. 1b y 1c) permite determinar los periodos diurno y semidiurno de 24 y 12 horas respectivamente (frecuencias $S_1$ (1 cpd[4]) y $S_2$ (2 cpd)) y el periodo de rotación solar junto con sus armónicas (27.00, 13.5 y 9.00 días), siendo la frecuencia $S_1$ la de mayor amplitud. En tanto que mediante el análisis de Fourier (figura 3), además de las frecuencias $S_1$ y $S_2$ se observa también las bandas laterales con frecuencias $S_{n,k} = (n + k/365.242199)$ cpd, con $n = 1,2,\ldots$; $k = \pm 1, \pm 2,\ldots$, generado por la frecuencia moduladora anual $h = 1/365.242199$ (que corresponde al periodo del año trópico) y sus armónicas sobre la variación solar diaria. Las bandas laterales son denotados por $S_{n,k}$, $n = 1,2,\ldots$; $k = \pm 1, \pm 2,\ldots$ (De Meyer, 2003). El efecto de la variación estacional es observado en la variación de la amplitud de H (figura 1a, 1d), son mayores en los meses de Marzo, Abril, Setiembre y Octubre, que corresponde a los dos equinoccios (mes de Lloyd E), similar variación que fue observado por Bartel y Jonhston (Bartels and Jonhston, 1940).

---

[1] Dip: distancia al Ecuador geomagnético.
[2] Meses de Lloyd: (D: Noviembre, Diciembre, Enero y Febrero, E: Marzo, Abril, Setiembre y Octubre, J: Mayo, Junio, Julio y Agosto).
[3] Sq: variación solar tranquila.

Para el análisis espectral de la variación L por wavelets es sustraídos las frecuencias que corresponden a la variación $S = S(t)$, para ello utilizamos la siguiente relación

$$S(t) = \sum_{n=1}^{4} \sum_{k=-2}^{2} A_{n,k} \cos(\omega_{n,k} t - \varphi_{n,k})$$

(*1*)

donde $\omega_{n,k} = 2\pi S_{n,k}$ y $S_{n,k} = 1/T_{n,k}$. $S_{n,k}$, $T_{n,k}$, $A_{n,k}$ y $\varphi_{n,k}$ son las frecuencias, periodos, amplitudes y fases respectivamente. Si $k = 0$ se tienen las variaciones principales y si $k \neq 0$ se tiene las variaciones de sus bandas laterales respecto a $S_n$.

Las frecuencias en el análisis espectral por wavelets para la variación lunar viene dado por la relación $L_n = (n + 2/M)$, donde $n = 1,2,\ldots$ y $M = 29.530588$ es el número de días solares en un mes lunar (mes sinódico lunar). Se muestra las frecuencias lunar diurna $L_1$ (0.9323 cpd) y semidiurna $L_2$ (1.9323 cpd), sus amplitudes son mucho menores incluso que el periodo de rotación solar y sus armónicas (figura 2b, 2c). El efecto de la variación estacional es observado en la variación de la amplitud de L, siendo entre los meses de Noviembre a Marzo mayor y el mes de Junio mínimo (figura 2d), lo cual también concuerda con los trabajos de Bartels y Johnston (1940).

---

[4]  cpd: ciclos por dia.

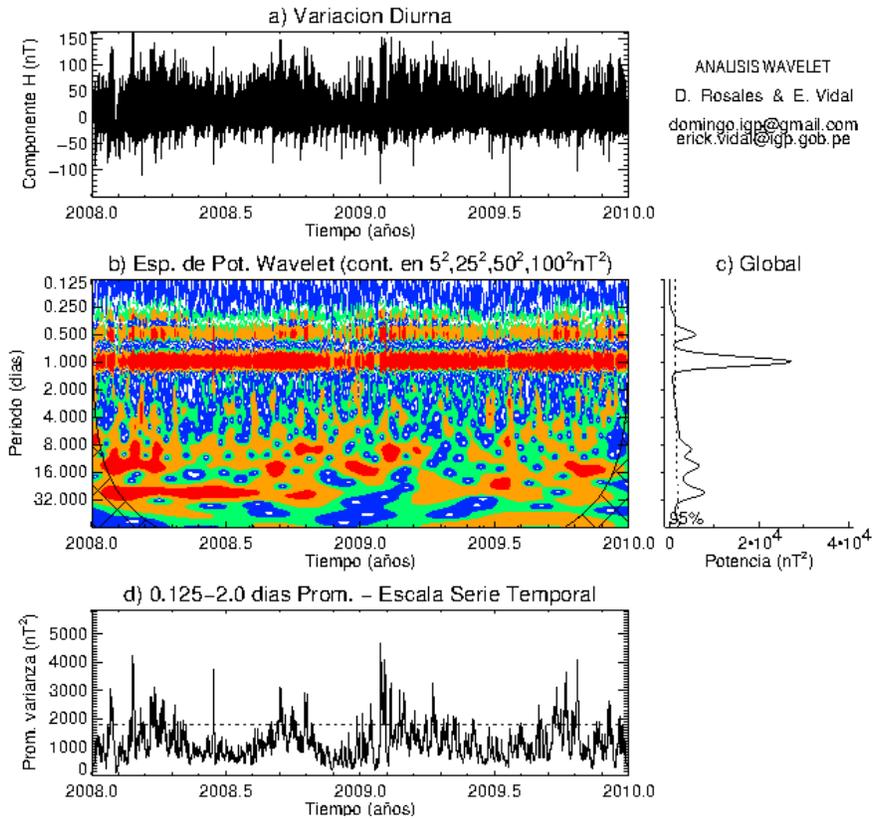

**Figura 1.**　(a) Intensidad horizontal H, (b) Espectro de potencia wavelet. Los niveles de contorno son en 25, 625, 2500 y 10000 $nT^2$. (c) Espectro de potencia wavelet global. (d) Varianza promedio entre 0.125 - 2.0 dias.

En el análisis espectral mediante la FFT para el efecto lunar (figura 3), además de las variaciones diurna $L_1$ y semidiurna $L_2$ muestra el efecto de sus bandas laterales debido al efecto del periodo del año trópico. Las frecuencias laterales son definido por la relación $L_{n,k} = (n + 2/M + k/365.242199)$ cpd, donde: $n = 1,2,...$; $k = \pm 1, \pm 2,...$ y $M = 29.530588$.

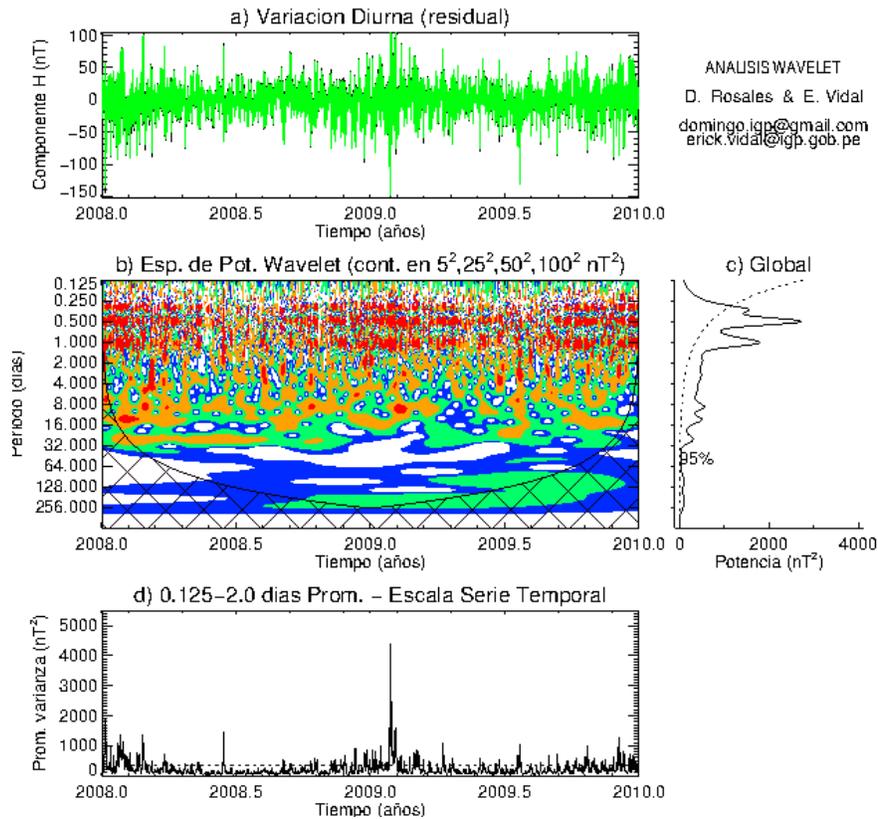

**Figura 2.**   (a) Intensidad horizontal H sin la variacion S, (b) Espectro de potencia wavelet. Los niveles de contorno son en 25, 625, 2500 y 10000 $nT^2$. (c) Espectro de potencia wavelet global. (d) Varianza promedio entre 0.125 - 2.0 dias.

**Conclusiones**

Fueron aplicados dos métodos de análisis espectral; wavelets y la transformada rápida de Fourier para determinar las frecuencias de la variación solar y lunar, presentes en los datos de la intensidad horizontal del observatorio de Huancayo para un periodo de dos años (2008-2009) comprendido dentro del rango de mínima actividad solar. Los análisis espectrales muestran una prominente amplitud en la variación diurna con frecuencia $S_1$ y sus bandas laterales $S_{1,-2}$, $S_{1,-1}$, $S_{1,1}$ y $S_{1,2}$. Así mismo la variación semidiurna es observable junto con sus bandas laterales ($S_2$ y sus bandas laterales $S_{2,-2}$, $S_{2,-1}$, $S_{2,1}$ y $S_{2,2}$) debido al efecto del periodo del año trópico (365.242199 días). La variación debido al efecto lunar también son observados en las variaciones diurna y semidiurna ( $L_1$ y $L_2$ ), juntamente con sus bandas laterales ( $L_{1,-2}$, $L_{1,-1}$, $L_{1,1}$, $L_{1,2}$, $L_{2,-2}$, $L_{2,-1}$, $L_{2,1}$ y $L_{2,2}$ ) debido al efecto del mes sinódico lunar (29.530588 días) y al año trópico.

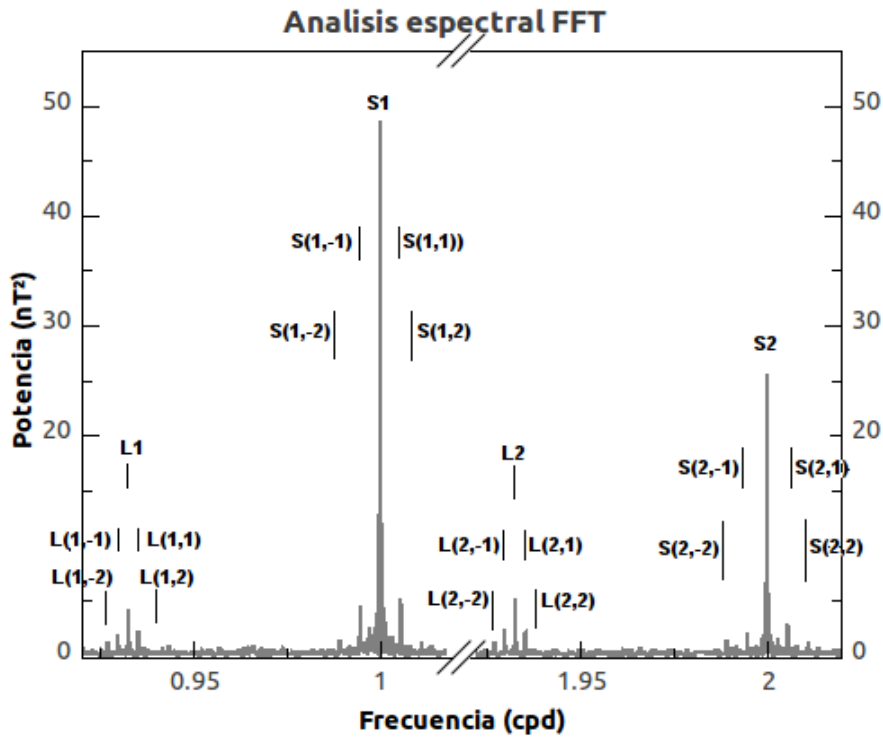

**Figura 3.** Espectro de potencia por medio de la transformada rapida de Fourier FFT para la variacion diurna y semidiurna solar y lunar.

En el análisis espectral por wavelets también verifica el periodo de 27.00 días y sus armónicas (13.50 y 9.00 días), periodos que corresponde al mecanismo de modulación debido a la variación de la actividad solar producido por el efecto de la rotación diferencial solar. La variación solar y lunar diaria exhiben variaciones estacionales con características distintas e independientes. La influencia del ciclo de actividad solar de 11 años no es observado en el presente análisis debido a que solo se ha tomado dos años de datos para el análisis espectral.

**Referencias**